# VARIABILITY OF ACCRETION DISKS SURROUNDING BLACK HOLES — THE ROLE OF INERTIAL-ACOUSTIC MODE INSTABILITIES

Xingming Chen[1] and Ronald E. Taam[2]

[1] Department of Astronomy and Astrophysics, Institute of Theoretical Physics, University of Göteborg and Chalmers University of Technology, 412 96 Göteborg, Sweden; e-mail address: chen@fy.chalmers.se

[2] Department of Physics and Astronomy, Northwestern University, Evanston, IL 60208; e-mail address: taam@ossenu.astro.nwu.edu

# ABSTRACT

The global nonlinear time-dependent evolution of the inertial-acoustic mode instability in accretion disks surrounding black holes has been investigated. The viscous stress is assumed to be proportional to the gas pressure only, i.e., $\tau = -\alpha p_g$. It is found that an oscillatory nonsteady behavior exists in the inner regions of disks ($r < 10 r_g$, where $r_g$ is the Schwarzschild radius) for sufficiently large $\alpha$ ($\gtrsim 0.2$), and for mass accretion rates less than about 0.3 times the Eddington value. The variations of the integrated bolometric luminosity from the disk, $\Delta L/L$, are less than 3%. A power spectrum analysis of these variations reveals a power spectrum which can be fit to a power law function of the frequency $P \propto f^{-\gamma}$, with index $\gamma \sim 1.4 - 2.3$ and a low frequency feature at about 4 Hz in one case. In addition, a narrow peak centered at a frequency corresponding to the maximum epicyclic frequency of the disk at $\sim 100 - 130$ Hz and its first harmonic is also seen. The low frequency modulations are remarkably similar to those observed in black hole candidate systems. The possible existence of a scattering corona in the inner region of the disk and/or to other processes contributing to the power at high frequencies in the inner region of the accretion disk may make the detection of the high frequency component difficult.

# 1. INTRODUCTION

The discovery of quasi-periodic oscillations (QPOs) in black hole candidate (BHC) systems provides a new probe by which the physical conditions and interactions in the environment surrounding a compact object can be investigated. Unlike the QPOs seen from the bright low mass X-ray binaries on the horizontal branch portion of the color-color diagram (see Lewin, van Paradijs, & van der Klis 1988; van der Klis 1989), QPOs from BHCs cannot be explained in the framework of the beat frequency modulated accretion model (Alpar & Shaham 1985; Lamb, Shibazaki, Alpar, & Shaham 1985) simply because of the absence of a solid surface and a magnetosphere surrounding the compact object. It is quite plausible that the BHC QPOs are produced in an accretion disk as a result of instabilities.

The properties which distinguish BHCs from other X-ray sources and the properties of these QPOs, in particular, have been recently summarized in the reviews by Lewin & Tanaka (1994) and by van der Klis (1994) respectively. QPOs with a wide range of frequencies have been observed from BHCs. For example, high frequency QPOs of about 6 Hz have been observed from GX 339–4 (e.g., Miyamoto et al. 1991) and QPOs of 3–10 Hz from Nova Muscae have been reported by Grebenev et al. (1991) and Dotani (1992). On the other hand, low frequency QPOs of about 0.04 Hz have been observed during the low state of Cyg X-1 (Angelini & White 1992; Vikhlinin et al. 1992a, 1994; Kouveliotou et al. 1992a), and GRO J0422+32 (Vikhlinin et al. 1992b, Kouveliotou et al. 1992b, Pietsch et al. 1993). Since the high and low frequency QPOs differ by a factor of a hundred, it is likely that different mechanisms are responsible for their production. For example, as shown by Chen & Taam (1994) the low frequency QPOs can be explained in terms of a thermal-viscous mode instability. On the other hand, the higher frequency QPOs may be related to an inertial-acoustic mode instability.

The inertial-acoustic mode, in its simplest form, is a sound wave modified by rotation. As applied to an axisymmetric, inviscid, non self-gravitating gaseous disk the effect of rotation on acoustic modes is described by the dispersion relation (e.g., Chandrasekhar 1961)

$$\omega^2 = \kappa^2 + k^2 c_s^2$$

for sound waves propagating in a direction perpendicular to the rotation axis. Here $\omega$ is the frequency of the perturbation, $c_s$ is the local sound speed, $k$ is the wave number in the radial direction, and $\kappa^2 = \frac{2\Omega}{r}\frac{d}{dr}(r^2\Omega)$, where $\kappa$ is the epicyclic frequency and $\Omega$ is the angular velocity at radius $r$. This dispersion relation illustrates the effect of the Coriolis force which produces a cutoff in frequency for long radial wavelengths. For a quasi-Keplerian disk, $\kappa \sim \Omega$, and solutions for $\omega$ are real and, hence, the oscillation mode is stable. For a given wave number $k$, $\omega$ has two real roots with opposite sign. These two solutions correspond to the inward and outward propagating waves.

In a viscous accretion disk, on the other hand, the above simple dispersion relation becomes a quartic equation and the inertial-acoustic mode can be either stable or unstable. The detailed local analyses have been undertaken by several groups (e.g., Kato 1978; Blumenthal, Yang, & Lin 1984; Wallinder 1991a,b; Chen & Taam 1993), and their studies show that the solutions for $\omega$ are, in general, complex. In the case that the corresponding modes are unstable, the instability takes the form of an overstable mode driven by viscous effects. The existence of this instability relies on the increase of the coefficient of viscosity upon compression in comparison with the expanded phase (see Kato 1978).

The local analyses, however, are insufficient to demonstrate that the accretion disk is unstable. Global studies must be carried out. For example, the nonlinear time dependent study of Chen & Taam (1992) demonstrated that the local instabilities can be stabilized due to propagation effects in accretion disks surrounding neutron stars in a manner similar to that described in an earlier work by Papaloizou & Stanley (1986) for disks surrounding white dwarf stars. On the other hand, it has been suggested that these inertial-acoustic mode instabilities may survive in the innermost regions of accretion disks surrounding black holes because of wave trapping in the inner region between the point corresponding to the location where the epicyclic frequency is maximum and the sonic point (Kato & Fukue 1980). In addition, because the sound speed declines rapidly toward the sonic point as a result of the effective advection of thermal energy into the black hole, the rate at which the perturbation escapes (approximately determined by the sound speed) can be smaller than the growth rate (Kato 1978). This has been confirmed by earlier nonlinear studies of Matsumoto,

Kato, & Honma (1988, 1989) and Honma, Matsumoto, & Kato (1992).

The oscillation time scale of the inertial-acoustic mode is comparable to the dynamical time scale, $t_d = r_g/c$, where $r_g = 2GM/c^2$ is the Schwarzschild radius of a black hole of mass $M$ and $c$ is the speed of light. Since $t_d \sim 10^{-4} - 10^{-3}$ s in the inner region of a disk surrounding a $10 M_\odot$ black hole, the observed frequencies (3–10 Hz) of the QPOs from BHCs are too low to be directly related to the oscillations of the inertial-acoustic mode itself. However, there are some suggestions from the isothermal disk calculation of Matsumoto, Kato, & Honma (1988) that the oscillation is modulated on a much longer time scale. Long term evolutionary calculations of the inertial-acoustic mode in disks surrounding black holes incorporating the full energy equation have not been carried out since thermal-viscous mode instabilities can occur which influence the time-dependent behavior of the accretion disk (Honma, Matsumoto & Kato 1991).

In this paper, we circumvent this difficulty and investigate the long term response of the accretion disk to these inertial-acoustic mode instabilities to determine their relevance to the observed QPOs at frequencies in the range of 3–10 Hz in BHCs. To suppress the thermal-viscous instability and to study the inertial-acoustic mode in isolation, we consider accretion disk models based upon a viscous stress which is proportional to the gas pressure only. It is well known that such disks are both viscously and thermally stable (Lightman & Eardley 1974). In the next section, we formulate the problem and outline the fundamental disk equations. In §3 we present the detailed numerical results of the global evolution for a number of sequences. The implications of our results for the interpretation of the high frequency QPOs from BHCs in terms of the inertial-acoustic mode in accretion disks will be discussed in the last section.

## 2. FORMULATION

We consider a model accretion disk surrounding a non-rotating black hole of mass $M$. The disk is assumed to be in a configuration which is axially symmetric, non self-gravitating, optically thick and geometrically thin. For such a thin disk, its structure and evolution can be described in terms of vertically integrated equations (see below). The gravitational potential of the central black hole is described approximately by the pseudo-Newtonian potential, $\Phi$, suggested by Paczynski & Wiita (1980), as $\Phi = -GM/(r - r_g)$ where $G$ is the gravitational constant and $r$ is the distance in the disk mid-plane from the black hole.

The structure of the disk is described by the surface density, $\Sigma$, the radial velocity, $v_r$, the angular velocity, $\Omega$, and the internal energy per unit mass, $\varepsilon_i$. In our approximation, these variables are functions of the radius, $r$, and the time, $t$, only. The hydrodynamical equations, corresponding to the conservation of mass, momentum, and energy, are cast in the conservative form. Specifically, the continuity of mass is written as

$$\frac{\partial \Sigma}{\partial t} + \frac{\partial}{r \partial r}(r \Sigma v_r) = 0. \tag{1}$$

The radial equation of motion is given by

$$\frac{\partial (\Sigma v_r)}{\partial t} + \frac{\partial}{r \partial r}(r \Sigma v_r^2) = -\frac{\partial P}{\partial r} + \Sigma(\Omega^2 - \Omega_k^2)r. \tag{2}$$

Conservation of angular momentum takes the form

$$\frac{\partial (\Sigma \Omega r^2)}{\partial t} + \frac{\partial}{r \partial r}(r \Sigma v_r \Omega r^2) = \frac{1}{r}\frac{\partial}{\partial r}\left(\Sigma \nu r^3 \frac{\partial \Omega}{\partial r}\right). \tag{3}$$

Finally, we write the energy equation as

$$\frac{\partial (\Sigma \varepsilon_i)}{\partial t} + \frac{\partial}{r \partial r}[r v_r (\Sigma \varepsilon_i + P)] = v_r \frac{\partial P}{\partial r} + \nu \Sigma \left(r \frac{\partial \Omega}{\partial r}\right)^2 - \frac{4 a c T^4}{3 \kappa \Sigma}. \tag{4}$$

These so called slim disk equations differ from the standard disk equations in that the radial velocity and departures from Keplerian rotation are explicitly taken into account (see also Abramowicz et al. 1988; Chen & Taam 1992, 1993). Here $T$, $P$ and $\nu$ are the mid-plane temperature, vertically integrated pressure, and viscosity respectively. The internal energy per unit mass, surface density, and temperature are related by the equation of state,

$$\varepsilon_i = [3(1 - \beta) + 1.5\beta] P/\Sigma, \tag{5}$$

and

$$P = P_g + P_r = \frac{\Sigma \mathcal{R} T}{\mu} + \frac{1}{3} 2 H a T^4. \tag{6}$$

Here $\mathcal{R}$, $a$ and $\mu$ are the gas constant, the radiation constant, and the mean molecular weight (assumed to be 0.62) respectively. Furthermore, the ratio of the gas to total pressure is defined as $\beta$, and the half thickness of the disk, $H$, as determined from the condition of hydrostatic equilibrium in the vertical direction is given as $H = c_s/\Omega_k$ where $c_s = \sqrt{P/\Sigma}$ is the local sound speed and $\Omega_k^2 = \frac{GM}{r(r-r_g)^2}$ is the Keplerian angular velocity.

The effective kinematic viscosity, $\nu$, is parameterized in terms of the standard $\alpha$ model pioneered by Shakura & Sunyaev (1973) and is prescribed to take the form

$$\nu = \frac{2}{3}\alpha\beta c_s H. \qquad (7)$$

Here, $\alpha \leq 1$ is taken to be a constant, independent of the local conditions in the disk. It is well known that disks based on this viscosity prescription are both thermally and viscously stable (see e.g., Lightman & Eardley 1974). Implicit in our study is the assumption that the viscosity can react on the time scale of the oscillation. This can occur for those turbulent eddies whose turn over time is less than the time of the oscillation (e.g., Goldreich & Keeley 1977) which is comparable to the local Keplerian time scale.

The above set of one-dimensional time-dependent hydrodynamical equations are solved with a code based on the explicit piece-wise parabolic method (Colella & Woodward 1984; Woodward & Colella 1984). The accretion disk is divided into $\sim 800$ grid points in the radial direction distributed equally on a logarithmic scale ranging from an inner boundary at $2.4\,r_g$ to an outer boundary at $300\,r_g$. The initial structure of the disk is taken from the transonic slim disk solutions constructed for this particular form of the viscous stress as in Chen & Taam (1993). Since the sonic point where the radial velocity is equal to the local speed of sound is usually located near a radius of $3\,r_g$, the inner boundary is situated in the supersonic region. At this boundary, it is assumed that the gradients of all dependent variables vanish. On the other hand, we assume that the dependent variables are fixed at their initial steady state values at the outer boundary. The structure of the disk is not particularly sensitive to the form of these boundary conditions.

## 3. NUMERICAL RESULTS

We have calculated the time dependent evolution of several model accretion disks for various viscosity parameters, $\alpha$, and mass accretion rates, $\dot{M}$, to investigate the global behavior of slim disks which are susceptible to the inertial-acoustic mode instabilities. The parameters characterizing each of the model sequences as well as the resultant behavior of such disks are summarized in Table 1. In all seven sequences, the mass of the black hole is chosen to be $10 M_\odot$. The mass accretion rate is measured in units of the Eddington limit defined as $\dot{M}_E = 64\pi GM/(\kappa_e c)$, where $\kappa_e$ is the electron scattering opacity.

Since the inertial-acoustic instability is related to the viscosity coefficient, we first focus on the dependence of the instability on the viscosity parameter, $\alpha$. The results from local stability analyses indicate that the growth rate of the instability scales approximately with the viscosity parameter $\alpha$ and that the instability is always present for nonvanishing $\alpha$. To investigate this dependence, we examined the time dependent behavior of the disk for $\alpha$ equal to 0.1, 0.2, and 0.3 in sequences 1, 2, and 3 respectively. For these three sequences the mass accretion rate was fixed at $\dot{M} = 0.1 \dot{M}_E$. The results of the evolutions indicate that the disk is globally time varying only for sufficiently large $\alpha$ ($\gtrsim 0.1$). This can be seen in Figure 1, where the time variation of the bolometric luminosity (obtained by spatial integration of the local surface flux over the entire disk) is illustrated for model sequences 1, 2, and 3 in panels a), b), and c) respectively. Specifically, the disk is stable for $\alpha = 0.1$ and nonsteady for $\alpha = 0.3$ with the transition between the two states occurring near $\alpha \sim 0.2$. It can also been seen that as $\alpha$ is increased from 0.2 to 0.3, the amplitude of the luminosity fluctuations ($\Delta L/L$) becomes more prominent increasing from 0.01% to 0.07%. This result is not expected from local analysis since such analyses predict that the disk is nonsteady for small $\alpha$ as well. This stabilization for $\alpha \lesssim 0.2$ is due to the global effects associated with wave propagation. Specifically, for small $\alpha$, the growth rate of the perturbation is not sufficiently large to overcome the escape rate (i.e., the rate at which the perturbation escapes), which is related to the local sound speed. That the disk is nonsteady for only large values of $\alpha$ has also been found by Matsumoto et al. (1989) and Honma et al. (1992). However, their lower bound for $\alpha$ lies between 0.05 and 0.1. This difference is due to the fact that the viscosity prescription used in these works is more temperature sensitive since it is based on a viscous stress which is proportional to the total pressure rather than the gas pressure only.

In order to investigate the mass accretion rate regime for which the disk is globally time dependent, we varied the mass accretion rate by a factor of 6 in sequences 4–7, keeping the viscosity parameter equal to unity. In Figure 2 the time variations of the disk luminosity are illustrated for sequences 4 and 5 for initial mass accretion rates of 0.2 $\dot{M}_{\rm E}$ and 0.3 $\dot{M}_{\rm E}$ respectively. It can be seen that the disk luminosity is nonsteady for sequence 4, but stable to short temporal variations in sequence 5 (see Figs. 2a, b). These results demonstrate that there is a tendency toward stabilization at higher mass accretion rates. In particular, the disk is globally stable for $\dot{M} \gtrsim 0.3\dot{M}_{\rm E}$. This result was not anticipated from the results of local analysis, which indicated that the instability was not dependent on $\dot{M}$ (see, e.g., Chen & Taam 1993). This result reflects the global effect related to the higher escape rate associated with the greater local sound speeds characteristic of disk models at higher mass accretion rates.

To illustrate the above stabilization effects due to a small viscosity parameter and to a high mass accretion rate we display the local sound speed as a function of radius in Fig. 3 for two choices of each disk parameter (sequences 1, 5, and 6). It is seen that, in the neighborhood of about $4 - 8r_g$, the sound speed scales with the mass accretion rate, but it is relatively insensitive to $\alpha$. Hence, by increasing $\dot{M}$ for fixed $\alpha$ (see sequences 5 and 6), the local sound speed increases, thereby increasing the escape rate of the perturbation. The tendency toward stability is indicated since the growth rate of the instability remains unchanged. Similarly, by decreasing $\alpha$ (see sequences 1 and 6), the growth rate decreases, but the speed of sound changes little. Hence, in this comparison, the growth rate decreases while the escape rate is not significantly affected. Thus, in either case, the escape rate may exceed the growth rate of the perturbation, and the disk tends toward a stable state.

The long term bolometric luminosity variations of a nonsteady accretion disk are shown for sequence 6 in Figure 4. This sequence is characterized by $\alpha = 1.0$ and $\dot{M} = 0.1\dot{M}_{\rm E}$. It can be seen that a range of time scales is manifested in the bolometric luminosity variations. Specifically, there are long term modulations on time scales $\sim$ 0.1 s as shown in panels a), b), and c) as well as short term modulations on time scales of $\sim 0.01$ s as shown in panel d). The amplitude of these variations is small, i.e., $\delta L/L \sim 1.5\%$; however, these amplitudes do not reflect the large changes in some of the local disk variables. In particular, the local mass accretion rate and the radial velocity can vary by a factor exceeding a hundred, a result which has also been obtained by previous workers (Honma et al. 1992).

The range in time scales present in the time series of the bolometric luminosity can be seen in the power spectrum of sequence 6 shown in Figure 5. In order to obtain a smoother power spectrum, the entire time series is broken into 5 - 9 equal segments which are overlapped by one half of their length (see Press et al. 1986). The power spectra are calculated for each of these segments and they are added to produce an average power spectrum. There is a significant peak centered at a frequency $\sim 115$ Hz corresponding closely to the maximum epicylic frequency in the disk and a very weak harmonic. The power spectrum is also seen to rise with decreasing frequency. Specifically, for frequencies between $\sim 4 - 40$ Hz the power spectrum can be fitted to a power-law function, $P \propto f^{-\gamma}$, for which $\gamma \sim 2.3$. In addition to the high frequency features, a weak low frequency feature can be seen in the power spectrum at a frequency $\sim 4$ Hz. The long term modulation can also be seen directly from the time series data (see Fig. 4).

To determine the generality of these results, we have also investigated a model accretion disk characterized by $\dot{M} = 0.05\dot{M}_{\rm E}$ and $\alpha = 1$ (see sequence 7). The time evolution of the bolometric luminosity emitted by the disk is presented in Figure 6 for a total evolution time of 6.5 s. It is quite similar to Figure 4; however, the amplitude of the variation of the luminosity is relatively larger, corresponding to about $\Delta L/L \sim 2.5\%$. This is because the sound speed is smaller for lower mass accretion rates so the escape rate of the perturbation is smaller and the disk, therefore, has a greater tendency toward instability. The power spectrum of this time series is illustrated in Figure 7. Beside the primary peak at about 135 Hz, the harmonic is also quite visible, which reflects the larger amplitude variations in the bolometric luminosity. It is seen that the high frequency component is a very weak function of the mass accretion rate. The slight difference in the high frequency component in sequences 6 and 7 may be caused by the effects of pressure which are more important at higher mass accretion rates where the angular frequency is lower. This power spectrum also rises with lower frequencies and can be fit to a power-law function, but for a wider range of frequencies than in model sequence 6. In this case, the power index is smaller and $\gamma \sim 1.4$. Although there are indications of long term modulations in light curve, there is no indication of a distinct low frequency feature in the power spectrum, in contrast to the case for model sequence 6.

## 4. DISCUSSION

We have investigated the global stability of the inertial-acoustic mode in accretion disks surrounding black holes. Nonlinear time dependent calculations demonstrate that the modes modify the disk luminosity by several percent and take the form of radial waves which propagate within the inner region of the disk. Specifically, for $r \lesssim 3.5 r_g$ there are indications that waves propagate inward, whereas for $r \gtrsim 3.5 r_g$ the waves propagate outward. This dichotomy can be attributed to the results of local analysis which reveal that for radii less than a critical radii the inward propagating mode is most unstable whereas for radii greater than this critical radii the outward propagating mode is more unstable (see Chen & Taam 1993). This critical radius is located near the maximum pressure in the disk in the steady transonic model and the change in the dominant mode is most likely related to the change in the sign of the pressure derivative at this point. To illustrate this property of the waves, the contours of the time-dependent local mass accretion rate with respect to radius and time are plotted in Fig. 8 for model sequence 6. We see no clear edge inside which waves are trapped. Instead, the unstable region becomes wider as the disk evolves. On the other hand, the kinetic energy (i.e., $\frac{\Sigma v_r^2}{2}$) is trapped in the inner regions of the disk. The size of this region changes insignificantly as the disk evolves. This can be seen in Figure 9 from the contours of the time dependent local kinetic energy at two different time intervals. Specifically, the outer edge of the region is at about 8 $r_g$.

There is also a similar kinetic energy trapped region for sequence 7. However, the local sound speed is reduced at lower mass accretion rates and there is more power at lower frequencies at about 1 Hz. We conjecture that the 4 Hz feature for sequence 6 is apparent because there is less power at 1 Hz, rather than more power at 4 Hz. On the other hand, there is more power at 1 Hz with about the same amount of power at 4 Hz in sequence 7 so the feature disappears.

The wavelength of the waves, can be inferred from Figures 8 and 9 and is between $1 - 1.5\,r_g$ in the regions between $3\,r_g$ and $8\,r_g$. This is larger than the thickness of the disk which is in the range between $0.01r$ to $0.1r$. This result is consistent with our one-dimensional approximation since for waves of wavelength less than the thickness of the disk the wave refracts to the disk surface (see Lin, Papaloizou, & Savonije 1990). The frequency of the waves can also be estimated from Figure 8 as $f \sim 1/\Delta t \sim 100$ Hz, which is near the peak frequency in the power spectrum. Here, $\Delta t$ is the time interval between crests of the waves.

The power spectrum of the light curve increases with decreasing frequency and is characterized by a power law in the frequency range 1 - 40 Hz with an index of $\sim 1.4 - 2.3$. In addition to the high frequency oscillations, the power spectrum reveals the presence of lower frequency modulations at 4 Hz, at least for one case. These longer timescale variations ($\sim 0.1$ s) are very reminiscent of the 3 - 10 Hz QPOs observed in BHCs. As found in the previous section, its appearance is very sensitive to the mass accretion rate and the viscosity parameter $\alpha$. These longer time scale modulations are similar in kind to that found in the isothermal disk calculation by Matsumoto, Kato, & Honma (1988) and are likely related to the time by which the inertial acoustic wave traverses the trapped region of the disk.

Although the 3 - 10 Hz feature is common in many BHCs, the feature at $\gtrsim 100$ Hz has not yet been observed. Most of the power spectra from observations are limited to less than 100 Hz (see van der Klis 1994). In the case of Cyg X-1 during the low state, power spectra extend beyond 100 Hz. If the long term modulations of the inertial acoustic modes are to be viable for the 3 - 10 Hz features and more black hole candidates are observed not to exhibit the high frequency feature, then some physical process must exist to reduce the amplitude of these high frequency oscillations and/or increase the noise level at high frequencies.

A possible solution is to invoke an electron scattering cloud or corona which envelopes the inner region of the disk. Provided that the optical depth of the surrounding corona is nonnegligible, the amplitude of the oscillations produced either by a rotating beam of X-rays or by fluctuations of the central source itself can be suppressed (see Kylafis & Klimis 1987). The former case may be responsible for the absence of the periodic oscillations (due to the beaming) at the neutron star rotation frequency ($\sim 100$ Hz) in the bright X-ray sources where only horizontal branch QPOs of frequency $\lesssim 50$ Hz are observed (see also Brainerd & Lamb 1987). In our case, however, a rotating beam is not involved as the source luminosity is fluctuating. Since the fluctuation is not purely periodic and the geometrical structure of the emitting region is disk shaped instead of a point source, the analytical solution of Kylafis & Klimis (1987) cannot be applied exactly. However, the size, $R$, and the optical depth to electron scattering, $\tau_{es}$, of the surrounding corona can be roughly estimated. In particular, for the amplitude of the source fluctuations of frequency $f$ to be significantly reduced, the photon escape time must be much longer than the timescale of the variability or $R\tau_{es} 2\pi f/c >> 1$. The amplitude of the observed oscillations will, then, be reduced by a factor $\delta$ where $\delta = \pi c/(6 f R \tau_{es})$. For $M/M_\odot = 10$ and $f \sim 100$ Hz, $\delta \sim 50/[(R/r_g)\tau_{es}]$. Results of studies based upon spectral fitting of unsaturated Comptonization models in BHCs indicate that $\tau_{es}$ is about 2-3 (see, e.g., Grabelsky et al. 1994). Hence, for $\delta < 0.05$, which may effectively smear out the high frequency oscillations, the size of the corona is required to be $\sim 300 - 500\, r_g$. The properties of this corona will not significantly affect the QPOs at 4 Hz since then $R\tau 2\pi f/c$ is less than unity. Although the size of the required corona is large, the presence of other processes which produces noise at high frequency may relax this constraint. For example, the observed power law of the noise for LMX X-3, LMC X-1, GS 2000+25, and GS 1124-68 in the high state is $P \propto f^{-1}$ (Ebisawa et al. 1989, Treves et al. 1990, Miyamoto et al 1992) suggesting that other processes are operating to increase the noise level above that calculated in our framework which will allow for a smaller corona.

If such a scattering cloud is present, our results may be relevant to the QPOs ($\sim 3 - 10$ Hz) observed from BHCs. Since the local dynamical time scales linearly with the mass of the central object, we expect that the QPO frequency will scale approximately inversely to the mass of the central object. However, since the size of the trapped region depends on the mass accretion rate, the range of observed QPOs may reflect variations of not only the black hole mass but also the mass accretion rate. More detailed studies of the dependence of the QPOs on the mass accretion rate will be the subject of future investigations.

This research has been supported in part by NASA under grant NAGW-2526. We would like to thank Prof. S. Kato for discussions during the workshop on slim disks held in Göteborg, Sweden. RT would also like to thank Dr. Typhoon Lee for the warm hospitality that was extended to him during his visit to the Academia Sinica Institute of Astronomy and Astrophysics in Taipei.

# Table 1
## Model Sequences

| Sequence | $\alpha$ | $\dot{M}$ | Stability |
|---|---|---|---|
| 1............ | 0.1 | 0.1 | stable |
| 2............ | 0.2 | 0.1 | marginal |
| 3............ | 0.3 | 0.1 | unstable |
| 4............ | 1.0 | 0.2 | unstable |
| 5............ | 1.0 | 0.3 | stable |
| 6............ | 1.0 | 0.1 | unstable |
| 7............ | 1.0 | 0.05 | unstable |

# FIGURE CAPTIONS

**Figure 1.** The time variations of the bolometric disk luminosity (in units of the steady state value) for model sequences 1, 2, and 3 are displayed in panels a, b, and c respectively. The model sequences are characterized by the same initial steady state mass accretion rate $\dot{M} = 0.1\dot{M}_{\rm E}$ but different viscosity parameters. Note that only for $\alpha \gtrsim 0.2$ is the disk unstable.

**Figure 2.** The time variations of the disk luminosity in units of the steady state value. Panels a and b correspond to model sequences 4 and 5 respectively. Here $\alpha$ is chosen to be unity. It is seen that for a mass accretion rate of $\dot{M} = 0.2\dot{M}_{\rm E}$ the disk is unstable whereas for $\dot{M} = 0.3\dot{M}_{\rm E}$ the disk is stable.

**Figure 3.** The sound speed (in the unit of cm/s) as a function of radial distance from a black hole for steady state disk models corresponding to sequence 1 (dashed), 5 (dotted), and 6 (solid). Note that the sound speed is much more sensitive to variations in $\dot{M}$ than $\alpha$. This sensitivity provides an explanation for the tendency for stability in disks at small $\alpha$ or/and at high $\dot{M}$.

**Figure 4.** The time variations of the disk luminosity (in the unit of the steady state value) for model sequence 6 with $\alpha = 1.0$ and $\dot{M} = 0.1$. It is seen that a range in timescales are present.

**Figure 5.** The power spectrum of the disk luminosity of model sequence 6. Besides the significant peak at $\sim 115$ Hz, a broad peak is evident near 4 Hz. In addition, the power increases toward lower frequencies as a power law with an index of 2.3.

**Figure 6.** The time variations of the disk luminosity of model sequence 7 with $\alpha = 1.0$ and $\dot{M} = 0.05$. It is seen that the amplitude of the oscillations is larger than that of model sequence 6.

**Figure 7.** The power spectrum of the disk luminosity of model sequence 7. The power spectrum can be fit to a power-law function of the frequency with index about 1.4. There is a narrow peak centered at $\sim 135$ Hz and also a strong harmonic.

**Figure 8.** Contours of the time-dependent local mass accretion rate with respect to the radius and time for model sequence 6. It is seen that there are waves moving out

from $\sim 3.5 r_g$ and of waves moving inward inside $3.5 r_g$.

**Figure 9.** Contours of the time dependent local kinetic energy with respect to the radius and time for model sequence 6. It is seen that the kinetic energy is trapped in the inner regions of the disk.